\documentclass[]{interact}

\usepackage{epstopdf}
\usepackage[caption=false]{subfig}

\usepackage{url}
\usepackage{algorithm}%
\usepackage{algorithmicx}%
\usepackage{algpseudocode}%
\usepackage{listings}%
\newcommand\StateX{\Statex\hspace{\algorithmicindent}}

\usepackage[numbers,sort&compress]{natbib}
\bibpunct[, ]{[}{]}{,}{n}{,}{,}
\renewcommand\bibfont{\fontsize{10}{12}\selectfont}

\newcommand{\piij}{\pi(i \textrm{ beats } j)}
\newcommand{\by}{\boldsymbol{y}}
\newcommand{\bt}{\boldsymbol{t}}
\newcommand{\blambda}{\boldsymbol{\lambda}}

\newcommand{\bx}{\boldsymbol{x}}
\newcommand{\boldzero}{\boldsymbol{0}}

\newcommand{\bz}{\boldsymbol{z}}
\newcommand{\bzeta}{\boldsymbol{\zeta}}

\begin{document}

\articletype{}

\title{Scalable Bayesian Inference for Bradley--Terry Models with Ties: An Application to Honour Based Abuse}

\author{
\name{Rowland G Seymour\textsuperscript{a}\thanks{CONTACT R.~G. Seymour. Email: r.g.seymour@bham.ac.uk} and Fabian Hernandez\textsuperscript{b}}
\affil{\textsuperscript{a}School of Mathematics, University of Birmingham, UK; \textsuperscript{b}Independent researcher, Costa Rica}
}

\maketitle

\begin{abstract}
Honour based abuse covers a wide range of family abuse including female genital mutilation and forced marriage. Safeguarding professionals need to identify where abuses are happening in their local community to best support those at risk of these crimes and take preventative action. However, there is little local data about these kinds of crime. To tackle this problem, we ran comparative judgement surveys to map abuses at local level, where participants where shown pairs of wards and asked which had a higher rate of honour based abuse. In previous comparative judgement studies, participants reported fatigue associated with comparisons between areas with similar levels of abuse. Allowing for tied comparisons reduces fatigue, but increase the computational complexity when fitting the model.  We designed an efficient Markov Chain Monte Carlo algorithm to fit a model with ties, allowing for a wide range of prior distributions on the model parameters. Working with South Yorkshire Police and Oxford Against Cutting, we mapped the risk of honour based abuse at community level in two counties in the UK. 
\end{abstract}


\section{Introduction}
Comparative judgement models have been to used to analyse a wide range of problems from predicting the outcomes of football leagues \cite{Cattelan2012} to student assessment \cite{Gray2024}. The aim of a comparative judgement study is to estimate the qualities of a set of objects. In a comparative judgement study, participants are shown pairs of objects and asked which meets object more strongly aligns with the study question. By allowing study participants to compare objects in terms of the quantity of interest rather than give absolute assessments of that quantity, comparative judgement models can elicit reliable and informative responses. In \cite{Jones2023}, the authors give the example of participants finding it easier to decide which of two circles has the larger area than estimating the size of an individual circle. The Bradley–Terry (BT) model \cite{Brad52} is  perhaps the most common model for analysing comparative judgement data. It used to rank and estimate features of a set of objects based on pair-wise comparisons. It assigns a quality parameter to each object in the study and estimates the value of these parameters from pairwise comparisons.

We were motivated by a project to tackle honour based abuse in the UK and support the development of community level safeguarding strategies. The specific aim of our project was to map the prevalence of honour based abuse at local level in two counties in the UK.  Abuse to control family members in order to protect the family's honour or standing is called honour based abuse. Perpetrators of honour based abuse often perceive a victim to be putting the family's honour at risk, or to be bringing shame upon the family \cite{Aplin2019}. Examples include Female Genital Mutilation (FGM), a practice that involve partial or total removal of the external female genitalia for non-medical reasons (typically to evidence that the girl is a virgin on the wedding day, and as such is an honourable bride), or forced marriage, where individuals are coerced to marry someone who is perceived to be honourable. These abuse causes significant physical and emotional harm to the victim and can cause both immediate and long-term health issues.

To map the prevalence of abuse at community level, we considered analysing existing data. However, we found existing data to be insufficient. Currently, the UK Government only provides regional level data about forced marriage, and national level data about other types of honour based abuse \cite{FMU19}. This makes developing effective safeguarding strategies for specific communities difficult. Other local level data does exist, as those providing services to victims may collect their own data. However, victims may be supported by a number of independent services, for example, law enforcement, health professionals or women's refuges \cite[][$\S7$]{Aplin2019}. This means that no one service has the complete picture of where honour based abuse is happening. Accessing data from independent services would require a large number of data sharing agreements and the transfer of sensitive data about victims of abuse, which was not feasible. Thus, we decided to collect our data using a method that avoids collecting personally identifying information. 

We decided to carry out a comparative judgement study. A comparative judgement study works by showing study participants pairs of wards in the community and asking them to choose which ward has the higher rate of the abuse in question. Study participants, known as judges, should be people who are familiar with the community and abuse, however the wisdom of the crowd type approach means that judges do not need to have a perfect knowledge. Running a comparative judgement study is a viable and effective way to measure the rate of abuses at community level, without the need to collect personally identifying or sensitive data. Indeed, it has been previously used to map forced marriage at community level in Nottinghamshire \cite{Seymour23} and deprivation at sub-ward level in Dar es Salaam, Tanzania \cite{BSBT}.  However, in previous studies, judges reported fatigue and difficulty providing pairwise comparisons when the levels of abuse in pairs of wards were similar \cite{Seymour23}. In particular judges, found it difficult to repeatedly make comparisons about pairs of wards which both had low levels of abuse. This has also been found to be an issue when using adaptive comparative judgement study designs \cite{Jones2023}. To mitigate this issue in our study, we allowed the judges to say (i) the level of abuse in one ward was higher than the other, (ii) the level of abuse was similar -- or tied -- in both wards, or (iii) they wanted to skip that particular pair of wards. 

The statistical aim of our work was to develop a method to fit a variant of the BT model that allows tied comparisons. In particular, we wanted to develop a method to fit such a model in a fast way, because our project partners had limited computing resources. Additionally, we wanted our method to allow for prior spatial assumptions to be included in the model. This is because in both counties there were a limited number of judges with the expertise to take part in our study, and including prior spatial assumptions has been shown to produce accurate results with a limited number of judges \cite{BSBT, Seymour23}.

Variants of the BT model allowing for tied comparisons are well developed, with the model described in \cite{Rao67} commonly used. The outcome of each comparison is modelled by a multinomial distribution, with the events being a win, a loss, and a tie. Inference for models with ties is, however, limited. Inference methods for simple models allowing for tied comparisons are fast and efficient but require large amounts of data. Inference methods have been developed for more sophisticated models can reduce the amount of data that need to be collected, but are slow to and expensive to use. 

We developed a new representation of the tied model by introducing a series of P\'olya-Gamma latent variables. This allows for both sophisticated prior correlation between the model parameters and scalable inference. or computing maximum likelihood estimates, an iterative algorithm is developed in \cite{Rao67}. Maximum likelihood methods exist \cite{Hankin2020}, but do not allow for prior spatial correlation to be assumed. Two Markov Chain Monte Carlo methods exist for generating samples from posterior distributions relevant to this model. In \cite{Caron2012}, the authors introduce a series of exponentially distribution latent variables to perform scalable inference. However, this method requires independent prior distributions to be placed on the model parameters, and an improper prior distribution on the parameter determining the importance of ties. In contrast, in \cite{BSBT}, the authors propose a Metropolis-Hastings random walk algorithm that allows for a wide range of correlations between the model parameters, and a choice of any prior distribution on the tie parameter, but scales poorly with respect to the number of wards in the study.

\subsection{Data}
Working with two groups of safeguarding professionals in the UK, we collected community level data about honour based abuse. In the county of South Yorkshire, we worked with South Yorkshire Police's Honour Based Abuse Unit to collect data about FGM in the county. FGM had been chosen by the Unit as a priority area for development. The county of South Yorkshire is split into four boroughs (Barnsley, Doncaster, Rotherham and Sheffield) and each borough is further divided into wards. There are a total of 95 wards in the county. We designed an interface that showed study participants pairs of wards and asked them about the risk of FGM in the wards. Participants could respond in three ways: i) choose the ward they thought had the higher risk, ii) say the risk in the wards were about equal, or iii) skip the pairing. 

Participants were recruited by South Yorkshire Police, who run a FGM network for frontline services and safeguarding professionals in the county. Participants were emailed a link to the online survey tool and were asked to make 30 pairwise comparisons, but could exit the survey at any point. In total, we collected 877 comparisons, of which 122 (13.9\%) were ties, from 18 participants.

In the county of Oxfordshire, we worked with Oxford Against Cutting, a charity tackling FGM and honour based abuse in the county. We focused on the city of Oxford and town of Banbury in this study, and judges were shown pairs of wards in these two locations and asked which had a higher rate of honour based abuse. Judges could also say the wards had a similar level of abuse. We collected 766 comparisons, of which 199 (26.0\%) were ties. 

\begin{figure}
    \centering
    \includegraphics[width = 1.05\textwidth]{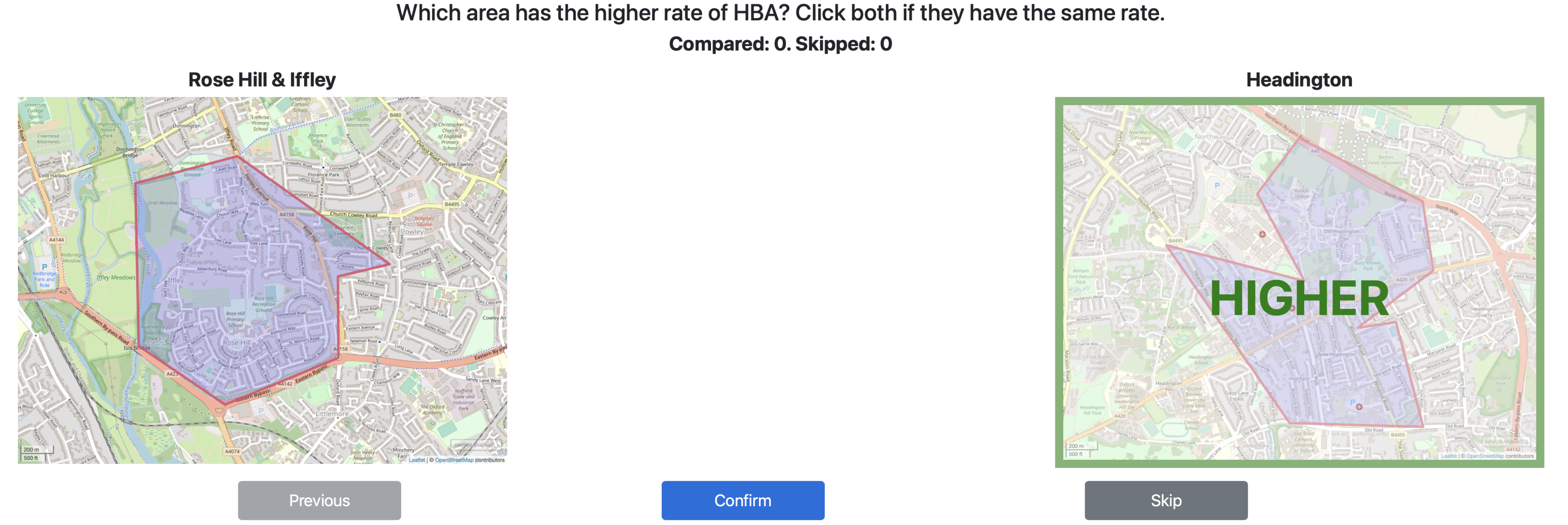}
    \caption{An example of the data collection interface we designed for this study. In this example of the Oxfordshire study, a judge was shown the wards Rose Hill \& Iffey and Headington. The judge chose Headington to have the higher rate of honour based abuse. The judge was also shown a counter of how many comparisons they had made.}
    \label{fig:interface}
\end{figure}

For both locations, data collection took place remotely though a bespoke website we designed. During the sign up process, judges were asked which wards they were able to make comparisons about (in South Yorkshire, these were the four boroughs, and in Oxfordshire these were Oxford and Banbury). Judges were only shown wards in the wards they were familiar and each pair of wards was chosen uniformly at random for the list of all pairs of wards the participant was familiar with. The judges was shown the name and maps of the wards in each pairing, as well as a counter of the number of comparisons they had made so far. An example of the interface is shown in Figure \ref{fig:interface}. Ethical approval was granted by the University of Birmingham's Science, Technology, Engineering and Mathematics Ethics Committee. 

\subsection{Structure of the article}
The remainder of the article is structured as follows. In Section \ref{sec: BT model}, we describe the standard BT model and the variant of the model that allows for tied comparisons. We then describe an MCMC algorithm for fitting the the model with tied comparisons to data in Section \ref{sec: Inference}. In Section \ref{sec: Sim studies}, we carry out simulation studies to determine the sensitivity of our model to the choice of hyperparameters and the scalability of the MCMC algorithm. In Section \ref{sec: results}, we use our MCMC algorithm to fit the model to analyse data we collected about honour based abuse in the counties of South Yorkshire and Oxfordshire, UK. In Section \ref{sec: conclusion}, we discuss our findings and method, and make suggestions about further work.

\section{The Bradley--Terry model} \label{sec: BT model}
\subsection{The Standard Bradley--Terry model}
One of the most widely used models for modelling comparative judgement data is the-so-called BT model \cite{Brad52} and is defined as follows. Consider a set of $N$ wards. We assign a parameter $\lambda_i \in \mathbb{R}$ to each ward describing its quality. Large positive values of $\lambda_i$ correspond to wards with low prevalence of honour based abuse and large negative values to wards with high prevalence of abuse. When comparing wards $i$ and $j$, the probability ward $i$ is judged to be superior to ward $j$ (for $i \neq j$) depends on the difference in quality parameters
\begin{equation}
    \textrm{logit}(\piij) = \lambda_i - \lambda_j \iff \piij = \frac{\exp(\lambda_i)}{\exp(\lambda_i) + \exp(\lambda_j)}  \label{eq: logit difference}
\end{equation} 
We can also write this probability in terms of the vector of regressors. Let $\boldsymbol{x}_{ij}$ be a vector of zeroes except the $i^{th}$  and $j^{th}$ entries, which take the values 1 and -1 respectively. By denoting the set of all ward quality parameters $\blambda$, we have $\bx^T_{ij}\blambda = \lambda_i - \lambda_j$ and the probability can be written as 
\begin{equation}
    \piij = \frac{\exp\left(\bx^T_{ij}\blambda \right)}{1 + \exp\left(\bx^T_{ij}\blambda \right)}.  \label{eq: regressors prob}
\end{equation}
We will use the formulation in Equation (\ref{eq: regressors prob}) in the construction of our MCMC algorithm. 

Under the assumption that comparisons between different pairs of wards and also comparisons for a given pair  are independent, the likelihood of the observed data is the following product of Binomials trials for each pair of wards. Let $n_{ij}$ be the number of times wards $i$ and $j$ are compared, $y_{ij}$ be the number of times ward $i$ is judged to be superior to ward $j$ and $\by$ the set of outcomes for all pairs of wards. The likelihood function can be written as
\begin{equation}
  \pi(\by|\blambda) = \prod_{i=1}^N\prod_{j < i} \begin{pmatrix} n_{ij} \\y_{ij} \end{pmatrix}\piij^{y_{ij}} (1-\piij)^{n_{ij} - y_{ij}}\,,  \label{eq: BT likelihood}
\end{equation}
where  $\boldsymbol{\lambda} = \{\lambda_1, \ldots, \lambda_N\}$ is the set of ward quality parameters. 

\subsection{The Bradley--Terry model with ties}
We can relax the assumption that comparisons are Bernoulli trials and allow for tied comparisons. Examples of tied comparisons include tied matches in sporting events or, in our motivating example, when judges cannot distinguish between the levels of honour based abuse in two wards. We follow \cite{Rao67} and assume the probability ward is $i$ judged to have a higher rate of abuse than ward $j$ and a tie between wards $i$ and $j$ (for $i \neq j$) are 
\begin{align}
    \piij &= \frac{\exp(\lambda_i)}{\exp(\lambda_i) + \exp(\lambda_j + \delta)}\label{eq: tie model prob winloss}  \\
    &= \frac{\exp\left(\bx^T_{ij}\blambda -\delta\right)}{1 + \exp\left(\bx^T_{ij}\blambda - \delta \right)}, 
    \\ \pi(i \textrm{ ties } j) &= \frac{(\exp(2\delta) - 1)\exp(\lambda_i + \lambda_j)} {(\exp(\lambda_i) + \exp(\lambda_j + \delta))(\exp(\lambda_i + \delta) + \exp(\lambda_j))}. \label{eq: tie model prob tie}  
\end{align}
The parameter $\delta \geq 0 $ contributes to the probability of a comparison resulting in a tie, with larger values of $\delta$ corresponding to a higher probability of a tie. 

Unlike the standard model, each comparison in this model has one of three outcomes. As such, we must now consider all permutations, not combinations, of wards. We denote the the number of times ward $i$ was judged to have a higher rate of abuse than ward $j$ by $y_{ij}$ and the number of ties by $t_{ij}$, and the corresponding vectors by $\by$ and $\boldsymbol{t}$. The likelihood function for this model is
\begin{align} 
    \pi(\by, \boldsymbol{t} \mid \blambda, \delta) &\propto \prod_{i=1}^N\prod_{j=1}^N  \piij^{y_{ij}} (1- \piij)^{y_{ji}} \pi(i \textrm{ ties } j)^{t_{ij}/2} \\
    & \propto (\exp(2\delta) - 1)^{\sum_{i=1}^N\sum_{j=1}^N {t_{ij}/2}} \prod_{i=1}^N\prod_{j=1}^N \pi_{ij}^{y_{ij} + t_{ij}} \\
    & \propto (\exp(2\delta) - 1)^{\sum_{i=1}^N\sum_{j=1}^N {t_{ij}/2}} \prod_{i=1}^N\prod_{j=1}^N \left(\frac{\exp(\bx_{ij}^T\blambda - \delta)}{(1+\exp(\bx_{ij}^T\blambda - \delta))}\right)^{y_{ij} + t_{ij}} \label{eq: BT ties likelihood}.
\end{align}

\section{Scalable Bayesian inference for the tied model} \label{sec: Inference}
By Bayes' theorem, the posterior distribution is given by
$$
\pi(\boldsymbol{\lambda}, \boldsymbol{t}, \delta \mid \boldsymbol{y}) \propto \pi(\by, \boldsymbol{t} \mid \blambda, \,\delta)\pi(\blambda)\pi(\delta). 
$$
An MCMC algorithm is needed to generate samples from this posterior distribution. As we are motivated by mapping honour-based abuse with local police forces and charities, it was important for us to develop an MCMC algorithm for the BT model with ties that was both allowed prior spatial correlation and was scalable. Prior spatial correlation in comparative judgement studies has been shown to considerably reduce the amount data that needs to be collected to obtain accurate estimates \cite{BSBT}, which us why we wished to include it. Scalability was important due to the computing resources of our partners. Computing resources for fitting models in local police forces and charities are extremely limited and running an MCMC algorithm for multiple hours is likely infeasible. Our partners in this project used laptops with small memory allocations, which needed to be used for other day-to-day tasks like online meetings and word processing. 

Currently, an MCMC algorithm does exist generate samples from the BT model with ties \cite{Caron2012}. This algorithm includes an exponentially distributed latent random variable, which allows the MCMC algorithm to scale well as the number of wards increases.  However, this algorithm requires the ward quality parameters to be \textit{a priori} independent. Given the strong spatial component to our application, we wished to assume spatial prior correlation on these parameters.  Additionally, the algorithm requires an improper prior distribution on the interval $(0, \infty)$ for the tie parameter $\delta$. This results in a full conditional distribution for $\delta$ that can be sampled from directly, but limits the inclusion of prior information about $\delta$

A Metropolis-Hastings random walk for the BT model without tied comparisons was developed in \cite{BSBT}. Although this does not allow for tied comparisons, it does allow to spatial prior correlation to be assumed for the ward quality parameters.  This method could be extended to include inference for a tie parameter $\delta$ in a straightforward manner. However, the algorithm scales very poorly with respect to the number of wards. In the example used in \cite{BSBT}, the authors use a Metropolis-Hastings random walk algorithm on a comparative judgement data set on deprivation in the 452 subwards of Dar es Salaam, Tanzania. Producing estimates for the quality parameters for all 452 subwards requires running the MCMC algorithm for 1,500,000 iterations. The Markov chains are slow to converge and their mixing is poor. Indeed, the first 500,000 iterations of the Markov chains are removed as a burn-in period, equating to around one hour of computer time. Extending this method to allow inference for a tie parameter would likely increase the computational cost of this method. 

As the inference for the BT model with ties has shown to be scalable when a latent variable is included \cite{Caron2012}, we decided to develop a latent variable representation of the BT model with ties, but ensure that the representation allows for correlated prior distributions and does not restrict the choice of prior distribution on $\delta$. For each pair of wards, we introduce a latent variable $z_{ij}$, which follows a P\'olya-Gamma (PG) distribution. This distribution has density equal in distribution to the sum of gamma random variables $g_k \sim \hbox{Gamma}(b, 1)$ given by
$$
\pi(z_{ij} \mid b, c) \overset{d}{=} \frac{1}{\pi^2}\sum_{k=1}^\infty \frac{g_k}{(k - 1/2)^2 + c^2/(4\pi)^2}, 
$$
where $c \in \mathbb{R}$. In \cite{Polson2013}, the authors show the PG distribution has the useful property
\begin{equation}
  \frac{(e^x)^a}{(1+e^x)^b} = 2^{-b}e^{(a - \frac{b}{2}) x}\int_0^\infty e^{-\frac{z x^2}{2}}\pi(z\mid b, 0)dz.   \label{eq: PG identity}
\end{equation}
As contributions to our likelihood function have the same functional form as the left hand side of Equation (\ref{eq: PG identity}), we will use this property to derive the latent variable augmented likelihood function. Consider the likelihood contribution from all comparisons, tied or otherwise, involving wards $i$ and $j$ (that is the $(i, j)^{th}$ term of the observed data likelihood function in Equation \ref{eq: BT ties likelihood}),
\[
\pi(y_{ij},\, t_{ij} \mid \blambda, \delta) \propto \left( \frac{\exp\left(\bx_{ij}^T \,\boldsymbol{\lambda} - \delta\right)}{\left(1 + \exp\left(\bx_{ij}^T \,\boldsymbol{\lambda} - \delta\right)\right)} \right)^{y_{ij} + t_{ij}}
\]
Using the property in Equation (\ref{eq: PG identity}), we can write this likelihood contribution as
\begin{align*}
    \pi(y_{ij},\, t_{ij} \mid \blambda, \delta)  \propto & \exp\left(\frac{1}{2}(y_{ij}- t_{ij})(\bx_{ij}^T\lambda-\delta) \,\right) \\
    & \times \int_{0}^{\infty}\exp\left(- z_{ij} \left(\bx_{ij}^T \,\boldsymbol{\lambda} - \delta\right)^2/2\right) \pi(z_{ij}|1, 0) \mbox{ d}z_{ij} \\
    \propto & \exp\left(\frac{1}{2}(\bx_{ij}^T \,\boldsymbol{\lambda} - \delta)\right) E_{\pi(z_{ij}|1, 0)}\left[\exp\left(- z_{ij} \left(\bx_{ij}^T \,\boldsymbol{\lambda} - \delta\right)^2/2\right)\right]. 
\end{align*}

The likelihood function can then be written as the product of this likelihood contribution over all pairs of wards, yielding
\begin{align*}
   \pi(\by, \boldsymbol{t}|\blambda, \bz, \delta) \propto &  \prod_{i = 1}^N \prod_{j = 1}^N  \pi(y_{ij},\, t_{ij} \mid \blambda, z_{ij}, \delta) \\ 
   \propto &  \,(\exp(2\delta) - 1)^{\sum_{i=1}^N\sum_{j=1}^N {t_{ij}/2}} \prod_{i = 1}^N \prod_{j = 1}^N  \Biggl( \exp\left(\frac{1}{2}(\bx_{ij}^T \,\boldsymbol{\lambda} - \delta)\right) \\
   & \times E_{\pi(z_{ij}|1, 0)}\left[\exp\left(- z_{ij} \left(\bx_{ij}^T \,\boldsymbol{\lambda} - \delta\right)^2/2\right)\right]\Biggl).   
\end{align*}

By Bayes' theorem, this posterior distribution including the latent variables $\bz$ is 
$$
\pi(\blambda, \bz, \delta \mid \by, \bt) \propto  \pi(\by, \boldsymbol{t}|\blambda, \bz, \delta)\pi(\blambda)\pi(\bz)\pi(\delta). 
$$
We will now show that when a joint prior distribution is placed on the ward parameter $\blambda$, the inclusion of the latent variables $z_{ij}$ allow us to derive a closed form for the full conditional distribution for $\blambda$. The result is a scalable MCMC algorithm for the BT model with ties including prior spatial correlation. 

To include prior correlation in the BT model, we follow \cite{BSBT} and place a multivariate normal distribution on the ward parameters $\blambda \sim N(\boldsymbol{\mu}, \Sigma)$. This allows for prior correlation to be assumed between the ward parameters and the choice of $\Sigma$ is a modelling choice to suit the application. In our application, we choose a network based spatial structure, where each of the wards are nodes and edges are placed between adjacent wards. The full conditional distribution for the ward parameters using this prior distribution is 
\begin{align*}
\pi(\blambda|\bz, \by, \bt, \delta) & \propto \pi(\blambda)  \prod_{i = 1}^N \prod_{j = 1}^N  \pi(y_{ij},\, t_{ij} \mid \blambda, \delta) \\  \\
& =  \pi(\blambda)\prod_{i=1}^{N}\prod_{j = 1}^N \exp\left(k_{ij}\, \left(\boldsymbol{x}_{ij}^T\,\boldsymbol{\lambda} - \delta\right) - z_{ij} \left(\boldsymbol{x}_{ij}^T \,\boldsymbol{\lambda} - \delta\right)^2/2 \right) \nonumber \\
& = \pi(\blambda) \prod_{i=1}^{N}\prod_{j = 1}^N\exp\left(\frac{z_{ij}}{2} \left(\boldsymbol{x}_{ij}^T \, \blambda - \delta - \frac{k_{ij}}{z_{ij}} \right)^2 \right),
\end{align*}
where $k_{ij} = y_{ij} - n_{ij}/2$. The last step follows by completing the square. We can treat the double product as matrix multiplication, and denoting  by $X$ the design matrix where each column is the vector of regressors $\bx_{ij}^T$, and by $\bzeta$ the vector containing $\delta - k_{ij}/z_{ij}$, for all pairs $i$ and $j$. The full conditional distribution resolves to 
\begin{align*}
& = \pi(\blambda) \exp\left(-\frac{1}{2}\left(\bzeta - X\blambda \right)^T Z (\bzeta - X \blambda) \right) \nonumber \\
& = \pi(\blambda) \exp \Bigg( - \frac{1}{2} \left(\blambda - X^{-1} \bzeta\right)^{T} \left( X^T Z X \right)  \left(\blambda- X^{-1}\bzeta \right)\Bigg), \nonumber 
\end{align*}
Therefore, the multivariate normal prior distribution on $\blambda$ induces conjugacy. Thus the full conditional distribution is 
\begin{equation}
\blambda \mid \by, \bt, \bz, \delta \sim N(\boldsymbol{\mu}, S) \label{eq:conditional_lambda}
\end{equation}
where $S = \left(X^{T} Z X + \Sigma^{-1} \right)^{-1}$,  $\boldsymbol{\mu} = S X^T \left((\by + \boldsymbol{t})/2 - \delta Z\boldsymbol{1})\right)$, and $Z = \hbox{diag}(z_{ij})$. 

The full conditional distribution for $z_{ij}$ is
$$
z_{ij} \mid \blambda, \by, \bt, \delta \sim PG\left(y_{ij} + t_{ij},\, \bx_{ij}^T\blambda - \delta\right).
$$
Sampling from this distribution can be efficiently done using an accept-reject algorithm based on the alternating-series method of \cite{Devroye2009}, and we use the function provided in the \texttt{BayesLogit} R package \cite{Polson2013}. 

The choice of prior distribution on the tie parameter $\delta$ is a modelling choice and our method allows for any non-negative prior distribution to be placed on $\delta$. For our motivation, we place an vague exponential prior distribution with rate $\chi$ on $\delta$. The full conditional distribution has no closed form and so we generate samples for $\delta$ using a Metropolis-Hastings random walk algorithm. The full conditional distribution is given by
\begin{align*}
   \pi(\delta \mid \by, \boldsymbol{t}, \blambda) \propto & (\exp(2\delta) - 1)^{\sum_{i=1}^N\sum_{j=1}^N {t_{ij}/2}} \\
   & \times \prod_{i=1}^N\prod_{j=1}^N \left(\frac{\exp(\bx_{ij}^T\blambda - \delta)}{(1+\exp(\bx_{ij}^T\blambda - \delta))}\right)^{y_{ij} + t_{ij}} \exp(\delta \chi). 
\end{align*}

\subsection{Markov chain Monte Carlo}
In order to sample from the posterior distribution, we develop a bespoke MCMC algorithm, which is shown in Algorithm \ref{alg: standard}. Model (\ref{eq: logit difference}) is invariant to translations and hence an identifiability constraint is necessary. To address this, at each iteration of the MCMC algorithm, we translate the ward qualities. This does not change the value of the likelihood function as it is invariant to translations. We adapt the translation proposed in \cite{Caron2012} and define
$$
\Lambda = \frac{1}{N}\sum_{i=1}^N \lambda_i
$$
to be the total of all the ward qualities and the normalised quality of ward $i$ respectively. Under the prior distribution $\blambda \sim N (\boldsymbol{\mu}, \Sigma)$ the distribution of $\Lambda$ is 
$$
\Lambda \sim N\left(\boldsymbol{1}^t\boldsymbol{\mu}, \frac{\boldsymbol{1}\Sigma\boldsymbol{1}^T}{N^2}\right),
$$
where $\boldsymbol{1} = (1, \ldots, 1)^T$ is a vector of ones.

\begin{algorithm}
\caption{The MCMC algorithm for BT model}
	\begin{algorithmic}[1]
		\State Initialise the chain with values $\lambda_1, \ldots, \lambda_N$ and $z_{1, 2} \ldots z_{N-1, N}$
		\StateX \textit{Repeat the following steps}
        \For{$\{i, j\} \in \{1,\ldots, N\}^2$}
            \State  Draw $z_{ij} \sim PG(1, \lambda_i - \lambda_j)$
        \EndFor
        \State Draw values for the model parameters $\blambda$
        \State Draw values for the tie parameter $\delta$
        \State Draw $\Lambda$
        \State Compute $\blambda = \blambda + \Lambda - \frac{1}{N}\sum_{i=1}^N\lambda_i$
	\end{algorithmic}
	\label{alg: standard}
\end{algorithm}

We have developed an R package that implements our MCMC algorithm. The \texttt{speedyBBT} R package is available at \url{https://github.com/rowlandseymour/speedyBBT}.

\section{Simulation studies} \label{sec: Sim studies}
To show the effectiveness and scalability of our model and latent variable MCMC algorithm, we ran several simulation study and sensitivity analyses. All simulations were carried out on a 2021 iMac with an M1 chip.

\subsection{Testing efficiency}
To compare our latent variable MCMC algorithm's efficiency to efficiently fit the BT model with ties to data against the Metropolis-Hastings random walk algorithm, we carried out a simulation study. We generated 25 synthetic data sets, including ties, on FGM abuse in South Yorkshire and sought to infer the abuse parameters using both MCMC algorithms. For each data set we draw a set of abuse parameters $\blambda$ from the prior distribution, simulate a set of comparisons and fit the model to comparisons infer the ward quality parameters. To construct the prior distribution, we incorporate spatial information, as proposed in \cite{BSBT}, by assuming that the levels of abuse in nearby wards had higher correlation than wards that are far away from each other. We construct a network representation of South Yorkshire, treating each ward as a node and placing edges between adjacent wards. To describe the correlation in terms of the network, we use the matrix exponential of the network adjacency matrix, $A$. This assigns high prior correlation to pairs of wards that are highly connected and low correlation to pairs of wards that are weakly connected. We construct the covariance matrix, by computing
\begin{equation}
    \Sigma  = \alpha^2 D^{-1/2}e^A D^{-1/2},  \label{eq: sigma normalising}
\end{equation}
where $\alpha^2$ is a signal variance parameter describing the overall strength of correlation in the prior covariance matrix, and $D$ is a diagonal matrix containing the elements on the main diagonal of $e^A$. Multiplying the matrix exponential on both sides by $D^{-1/2}$ has the effect of normalising the matrix, so we can describe the prior variance through $\alpha^2$. Small values of $\alpha^2$ correspond to applications with very strong spatial correlation and as $\alpha^2$ increase the prior correlation decreases in strength. The prior distribution is $\blambda \sim N(\boldzero, \Sigma)$. 

For each synthetic data set, we simulated 800 comparisons according to the model in Equations \ref{eq: tie model prob winloss} and \ref{eq: tie model prob tie}. This is inline with the number of comparisons we collected in South Yorkshire. We draw a value of the tie parameter $\delta \sim U[0, 1]$. We set the prior variance parameter $\alpha^2 = 1$.

\begin{figure}
    \centering
    \includegraphics[width = 0.49\textwidth]{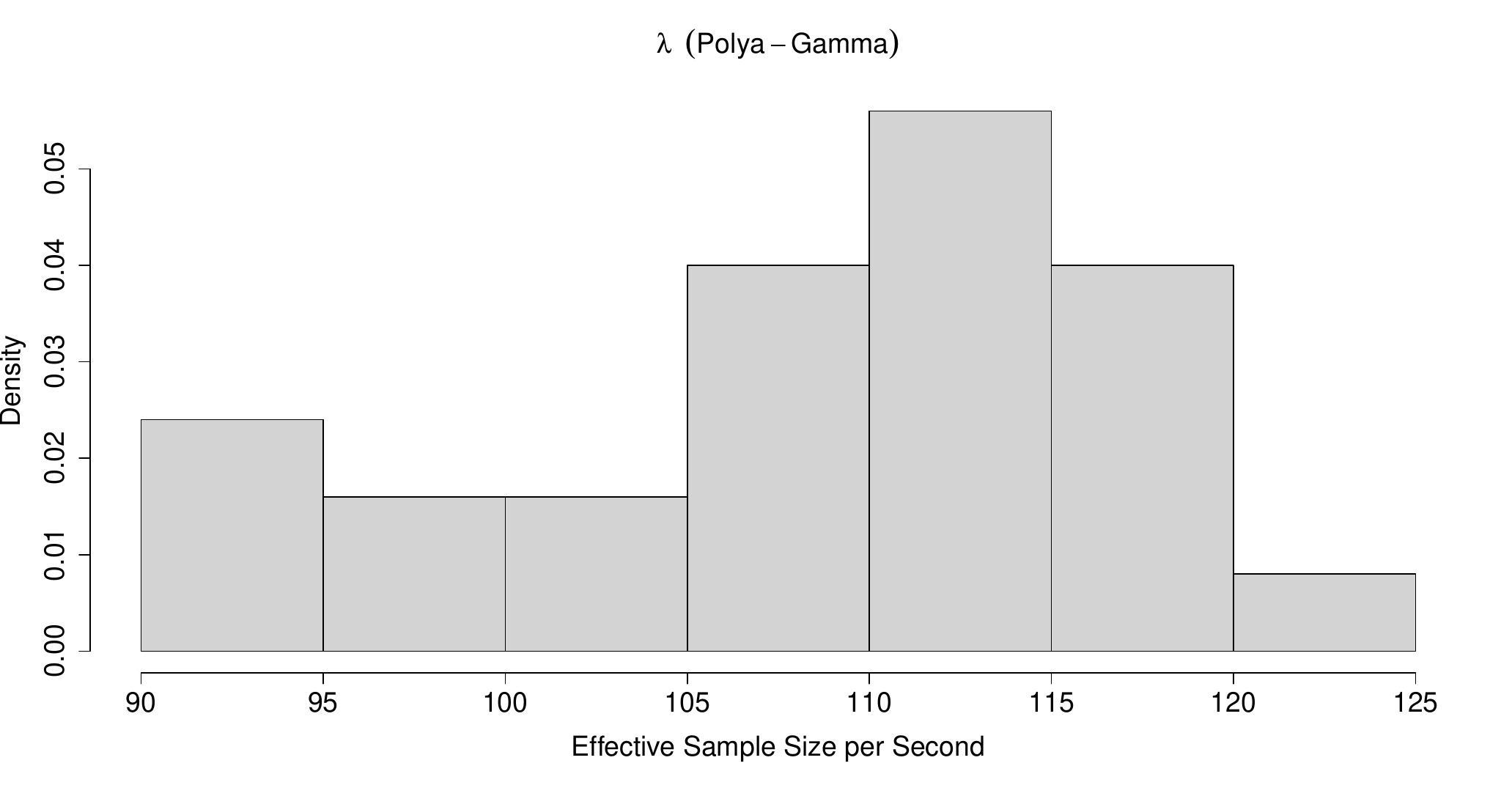}
    \includegraphics[width = 0.49\textwidth]{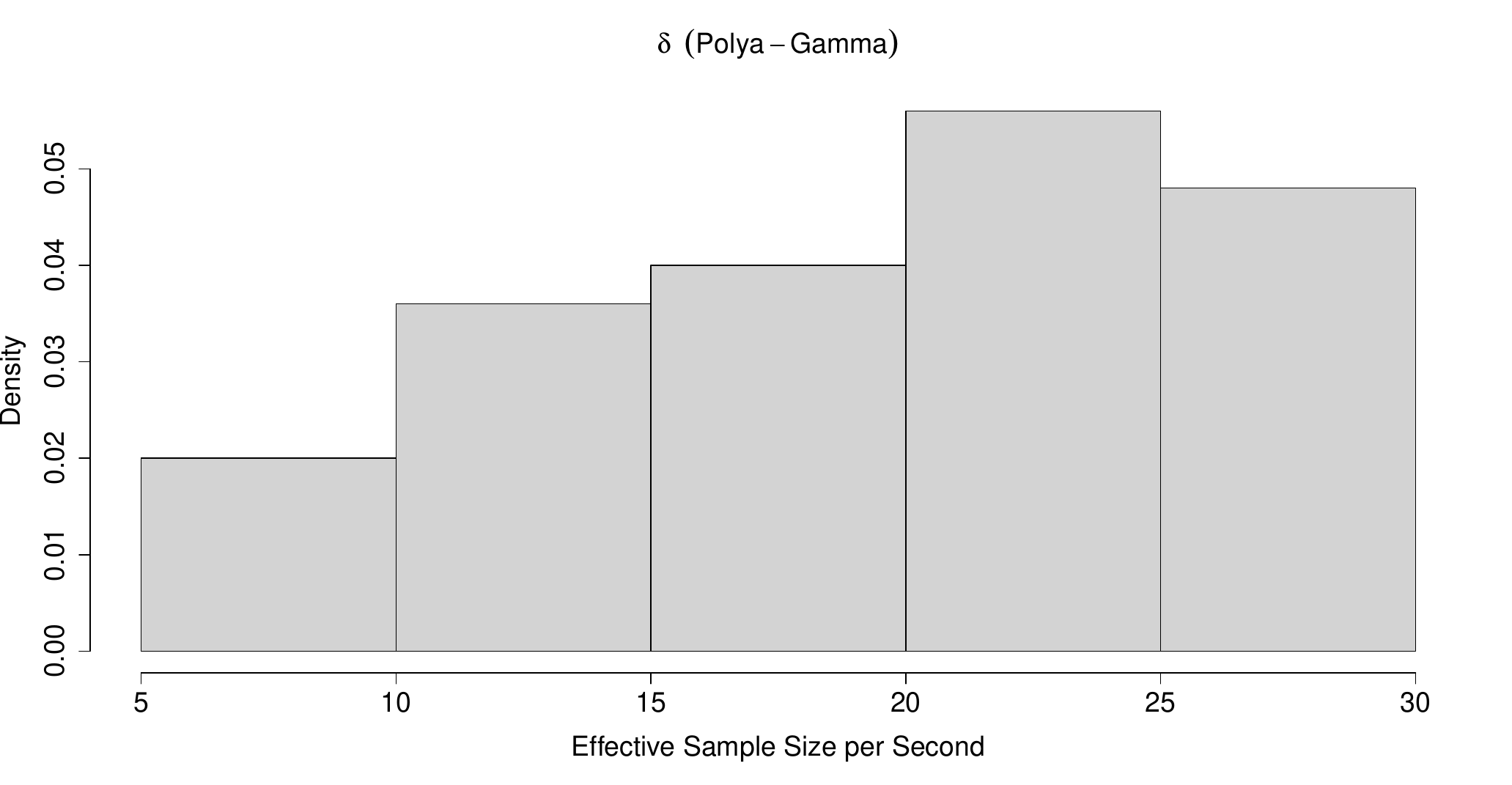}
    \includegraphics[width = 0.49\textwidth]{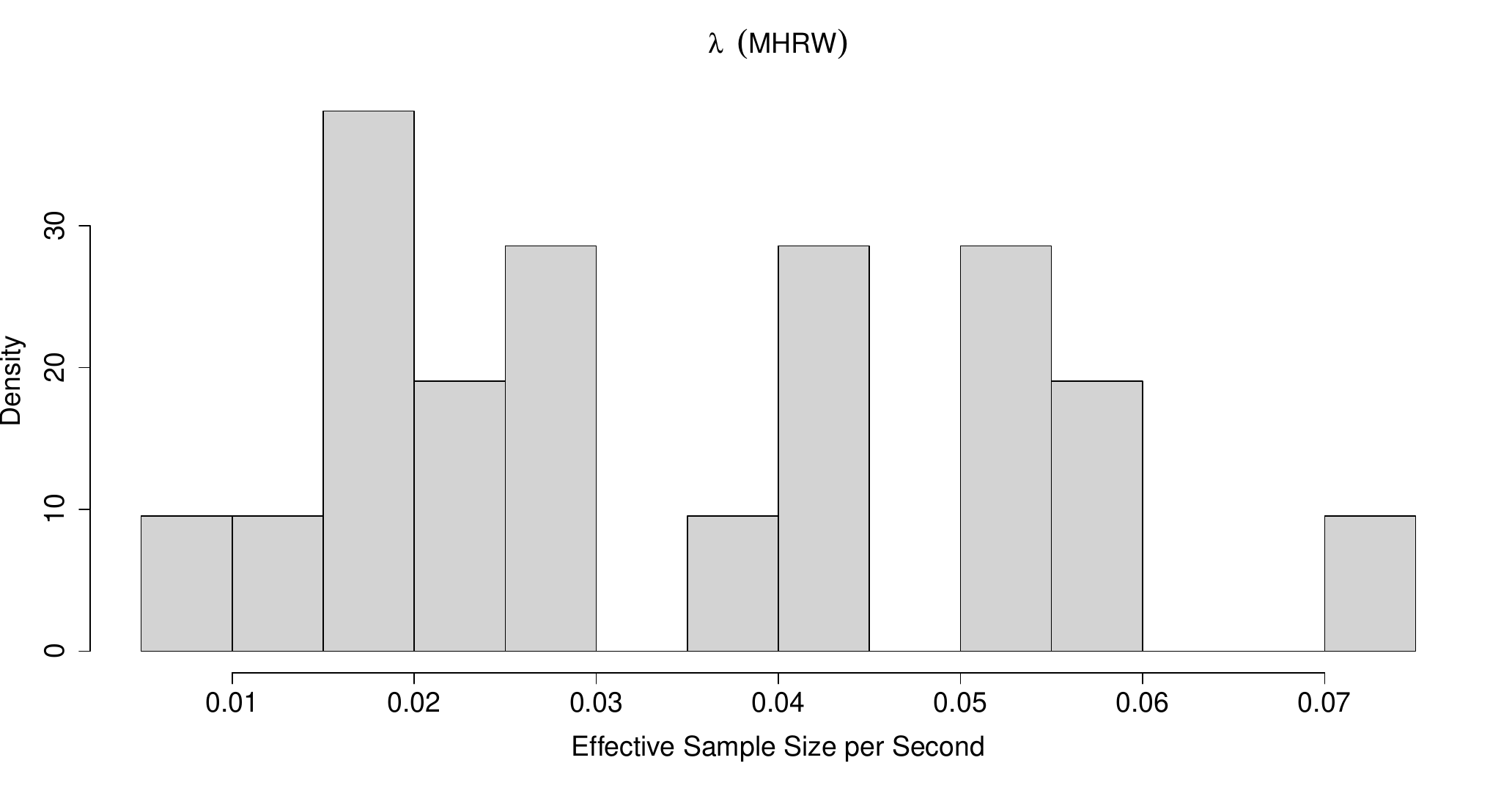}
    \includegraphics[width = 0.49\textwidth]{figures/ESS_Lambda_MHRW.pdf}
    \caption{Histograms showing the effective sample size per second for the simulation study. The top row shows the values for the P\'olya-Gamma latent variable representation and the bottom row the for Metropolis-Hastings random walk representation.}
    \label{fig: ESSs}
\end{figure}

For each of the simulated data sets, we recorded the effective sample size per second for the tie parameter $\delta$ and the average effective sample size per second across all risks parameters $\blambda$. Efficient samplers will have a higher effective sample size per second. Histograms of these values are shown in Figure \ref{fig: ESSs}. Overall, the simulation study shows the main advantage of the P\'olya-Gamma latent variable representation, which is a large improvement in fitting the Bradley--Terry model with ties to data, while allowing for correlated prior distributions on the parameters $\blambda$. Further details can be found in the Supplementary Material 

\subsection{Testing scalability}
To test the scalability of our MCMC algorithm, we ran a simulation study comparing it to a Metropolis-Hastings random walk algorithm. We generated synthetic data sets with 16, 32, 64, 128, 256, 512, and 1024 wards. For each data set we simulated a prior covariance matrix from a Wishart distribution with the identity matrix as the scale matrix and the number of wards as the degrees of freedom. We then normalised that matrix as shown in Equation (\ref{eq: sigma normalising}). We drew a set of ward quality parameters from a multivariate normal distribution with zero mean and covariance matrix given by the normalised synthetic prior covariance matrix. Finally, we generated a synthetic set of comparisons using the simulated ward quality parameters. The number of comparisons we generated was ten times the number of wards, as this is the recommended amount of data to collected in a comparative judgement study \cite{Jones2023} and close to the amount we collected in South Yorkshire.  We also fixed $\delta= 0.5$ when simulating the comparisons as this is similar to the value we inferred in the South Yorkshire data set. 

\begin{figure}
    \centering
    \includegraphics[width=0.8\linewidth]{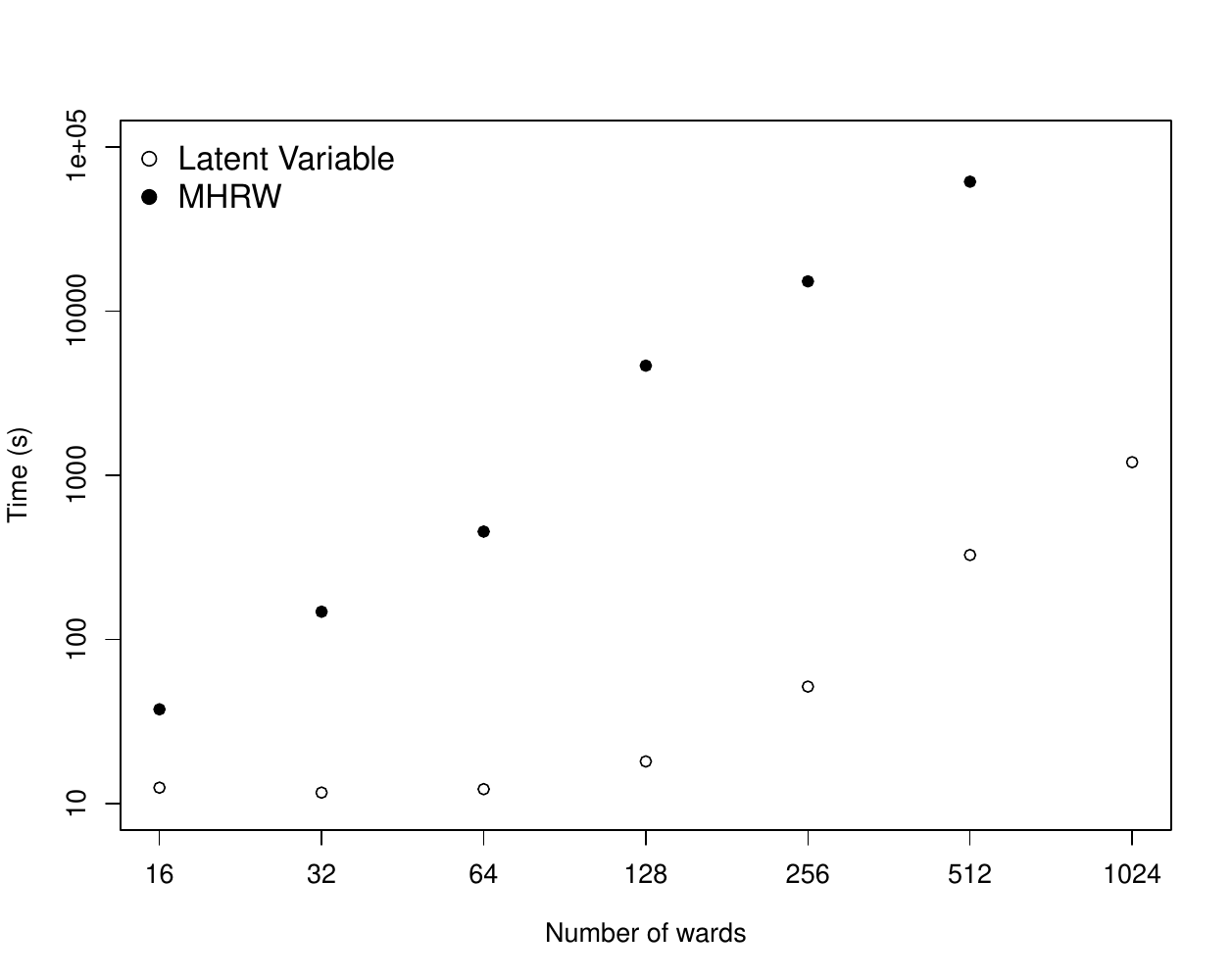}
    \caption{The mean time take to fit the the models to data sets of different sizes. The clear points correspond to the latent variable algorithm and the solid points to the Metropolis-Hastings random walk algorithm. Both axes are shown on a log scale.}
    \label{fig: time comparison}
\end{figure}

For each value of the number of wards, we generated ten synthetic data sets and fitted the model to each one using the latent variable algorithm and the Metropolis-Hastings algorithm described in \cite{BSBT}. We ran our latent variable algorithm for 5,000 iterations, treating the first 100 iterations as a burn-in period. The Metropolis-Hastings random walk algorithm took longer to converge, and we ran it for 100,000 iterations removing the first 1,000 as a burn-in period. We did not fit the Metropolis-Hastings random walk to the data sets with 1,024 wards as we expected each of the ten runs to take around 6 days. Figure \ref{fig: time comparison} shows the mean time taken to run the MCMC algorithms across each of the 10 runs. Our latent variable representation provides significant time advantages compared to the Metropolis-Hastings random walk algorithm and scales well as the number of wards is increased. With 1,024, our latent variable representation took on average 20 minutes to run. For data sets of the size we collected with our partners (64 - 128 wards), the model took around 20 to 40 seconds to run. This compares with 8 to 77 minutes using the Metropolis-Hastings random walk algorithm. 

\subsection{Sensitivity to the tie parameter and prior variance hyperparameter}
To determine if our model was sensitive to the value of the prior variance parameter $\alpha^2$ in Equation \ref{eq: sigma normalising} or the value of the tie parameter $\delta$, we carried out a simulation study for each parameter. To determine the sensitivity to the prior variance parameter, we fitted the model to the South Yorkshire data fixing the value of $\alpha^2$ to different values. These values represent, very strong, strong, and weak levels of prior spatial correlation, and the values were chosen based on the end points of the 95\% credible interval. We found that there was little difference in the estimates for the the ward quality parameters when using strong or weak prior spatial correlation to when inferring the parameter. Only when assuming very strong prior correlation (fixing $\alpha^2$ to a small value outside of the 95\% credible interval when it is learned) did we see a difference. In this simulation, there was substantial spatial smoothing and shrinkage of the ward quality parameters. Figures showing the effect of this smoothing are shown the Supplementary Material. 

To assess the sensitivity of our model to the tie parameter $\delta$, we simulated different data sets of FGM in South Yorkshire using different values of $\delta$. The values were chosen such that approximately 5\%, 20\%, 50\% and 75\% of the comparisons were tied. We found we could accurately recover the value of the tie and ward quality parameters, however there was a slight increase in the error when the data set comprises of 5\% and 75\% of ties. In the case where 5\% of the comparisons were tied, the model had less information to learn about the value of $\delta$ and made small adjustments to the values of the ward quality parameters to account for this. In contrast, when 75\% of the comparisons were tied, the model lacked enough information to accurately infer the values of the ward quality parameters. Given in the data we collected, around 15\% and 25\% of the comparisons are tied, we believe we are able to fully recover all model parameters. 

Full details of both of these analyses are presented in the Supplementary Material.

\section{Results} \label{sec: results}
\subsection{Female genital mutilation in South Yorkshire}
We collected 877 comparisons from 18 participants over a three week period in November and December 2022. There were 122 tied comparisons, that is the participants said the risk of FGM in these pairs of subwards was equal. We fitted the model to the data and ran the the MCMC algorithm for 5,000 iterations treating the first 100 iterations as a burn-in period. The prior distribution with spatial correlation for the ward quality parameters was given by network structure, as described in Section 4. For this prior distribution's variance parameter $\alpha^2$, we use a conjugate inverse-gamma prior distribution with shape and scale set to 0.01. For the tie parameter $\delta$, we use an Exp(0.01) prior distribution, as this is sufficiently vague. The posterior median estimates for risk of FGM in each ward and the values of the posterior variance are shown in Figure \ref{fig:SY Results}. Trace plots and convergence diagnoses can be found in the Supplementary Material.

\begin{figure} 
    \centering
    \includegraphics[width = 0.95\textwidth]{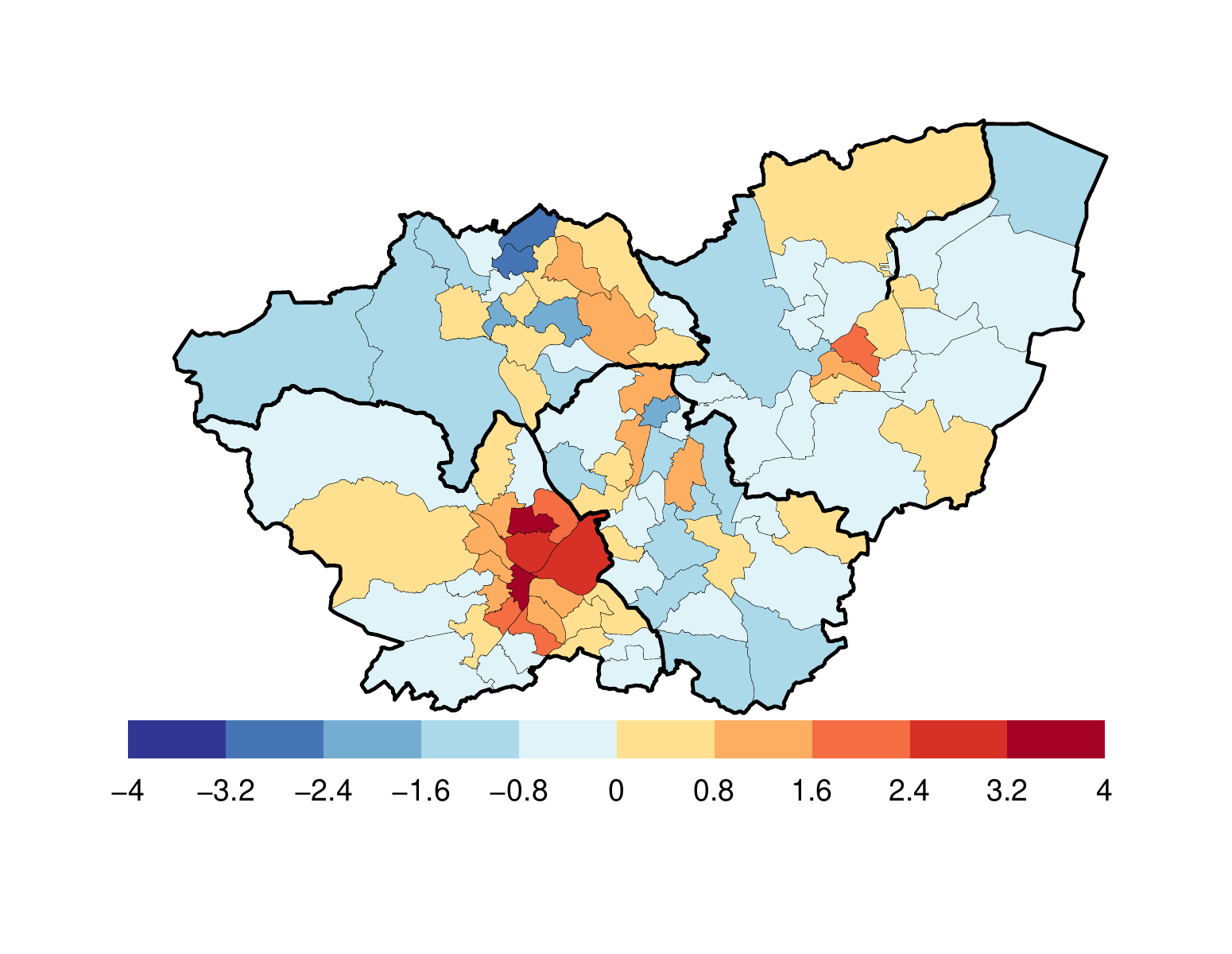}
    \includegraphics[width = 0.95\textwidth]{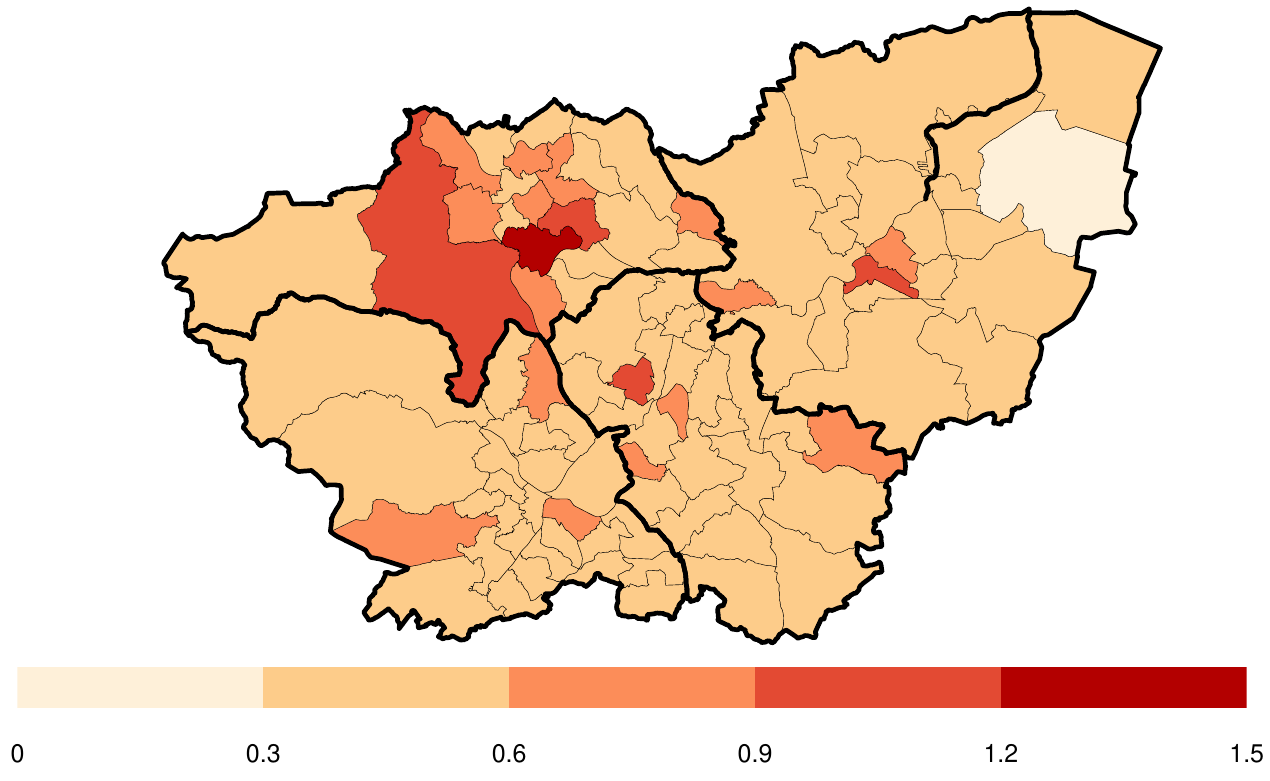}
    \caption{Top: Posterior median estimates for the relative risk of FGM in each ward. Blue wards have the lowest estimated risk and red wards the highest. Thick black lines show the outlines of the four boroughs. Bottom: Posterior variance for the estimates. Dark red wards have the highest risk.}
    \label{fig:SY Results}
\end{figure}

Our estimates show that the city of Sheffield, in the South West, has the highest risk of FGM in South Yorkshire. The highest risk wards are clustered around the centre of the city, with the four highest risk wards in the county all being in the city. In the borough of Doncaster (the eastern most borough), the risk is concentrated in Town ward, with the rest of the wards in the borough having medium to low risk of FGM. In Barnsley (the north-western borough), the risk of FGM is predominantly to the east of the borough, with the northern wards having the lowest risk in the whole of South Yorkshire. The uncertainty in our estimates is highest in Barnsley, as this was the borough where we recruited the fewest participants. 

The posterior median value for the tie parameter $\delta$ is 0.468 (95\% CI (0.390, 0.552)). Using the posterior median value for $\delta$ means that if a judge is shown a pair of wards with identical risks of FGM, the probability of a the judge declaring a tie is 0.139. This suggests the effect of ties were important in this study, not only reducing the risk of survey fatigue, but also providing us with information about risk of FGM in the city. 

\begin{figure}
    \centering
    \includegraphics[width = 0.49\textwidth]{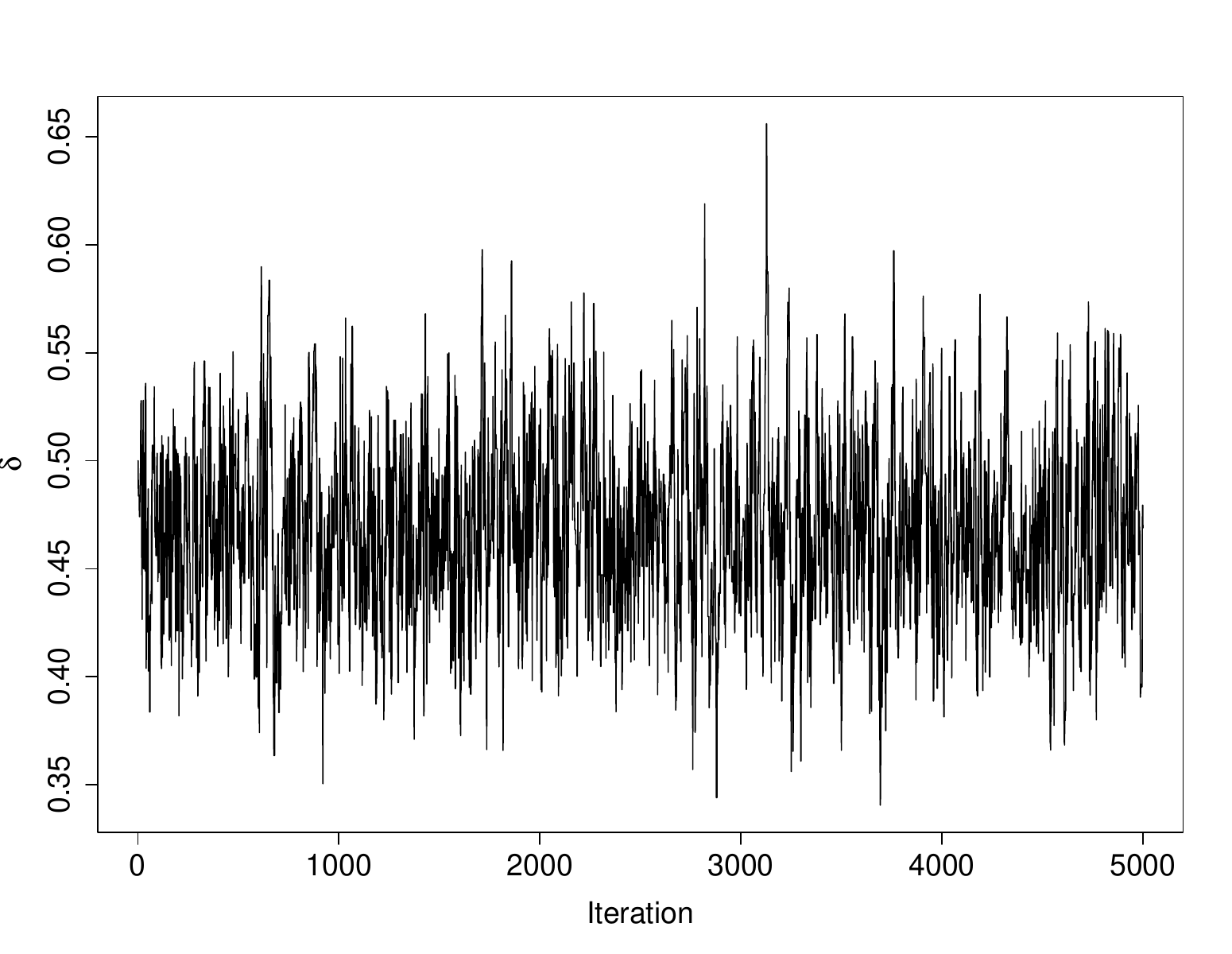}
    \includegraphics[width = 0.49\textwidth]{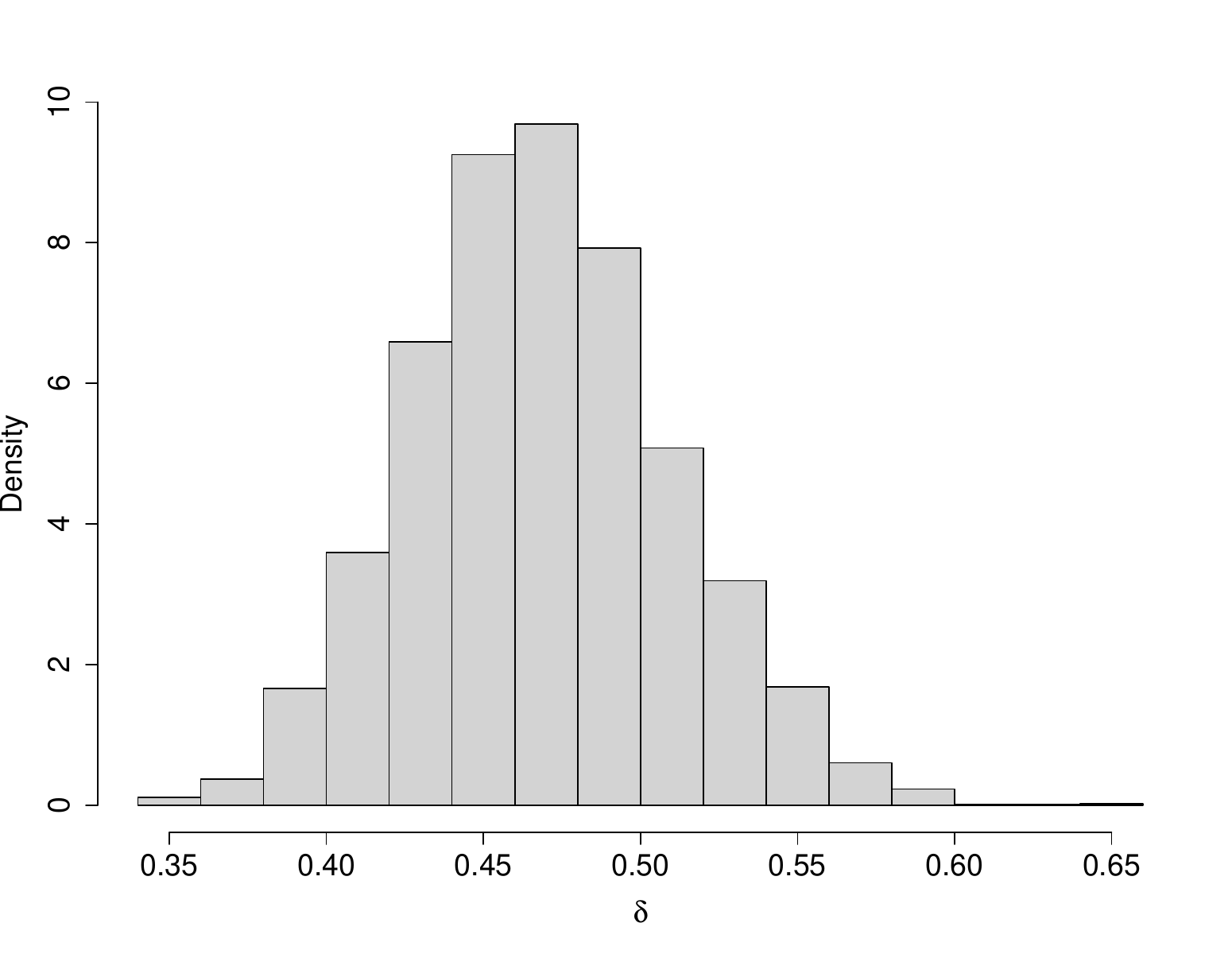}
    \caption{The trace plot (left) and histogram (right) of the posterior samples for the tie parameter $\delta$ in the South Yorkshire Study.}
    \label{fig:SY ties}
\end{figure}

\subsection{Honour based abuse in Oxfordshire}
We collected 766 comparisons from 12 participants in Autumn 2023. 199 of the comparisons were tied, representing 26.0\% of the data set. We fitted the model to the data and ran the MCMC algorithm for 5,000 iterations treating the first 100 iterations as a burn-in period. Trace plots and convergence diagnoses can be found in the Supplementary Material.

\begin{figure}
    \centering
    \includegraphics[width = 0.48\textwidth]{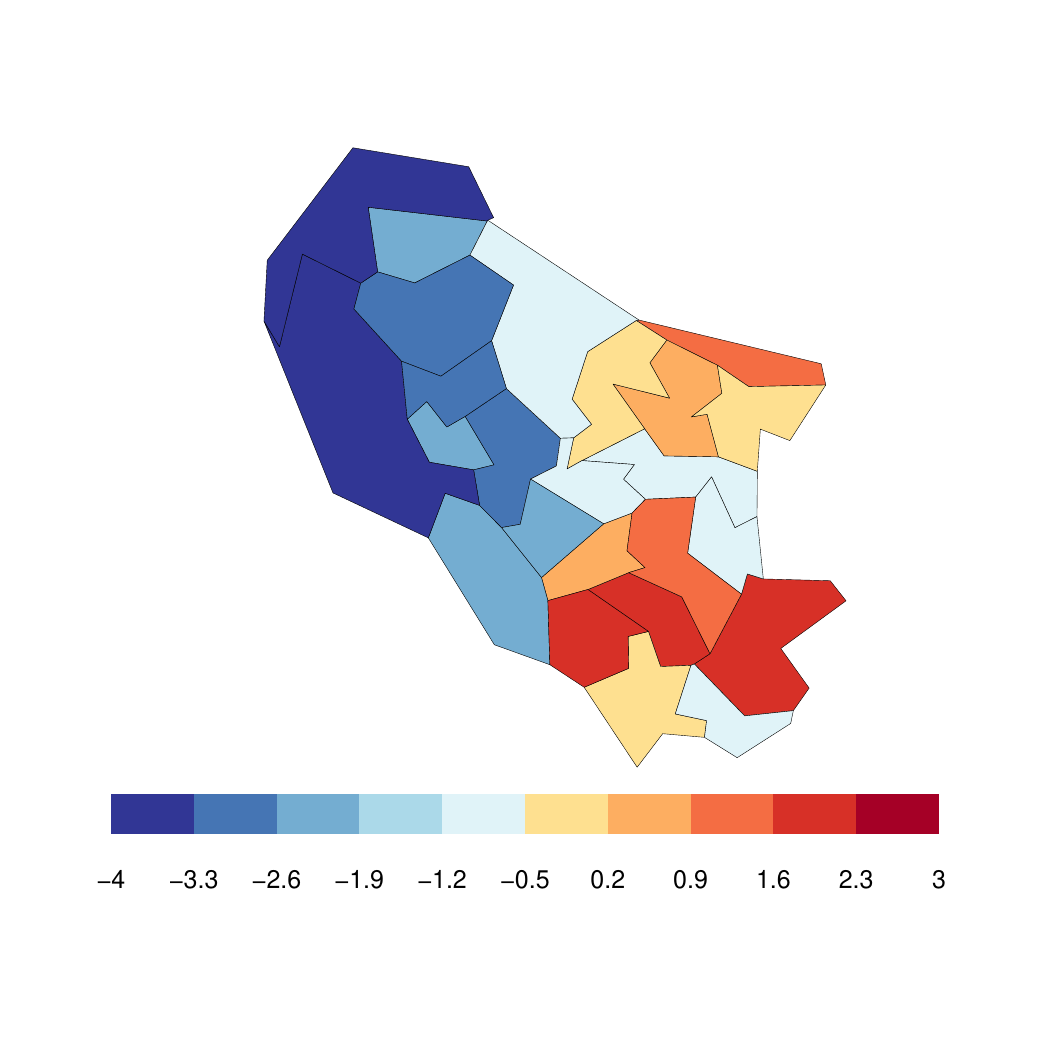}
    \includegraphics[width = 0.48\textwidth]{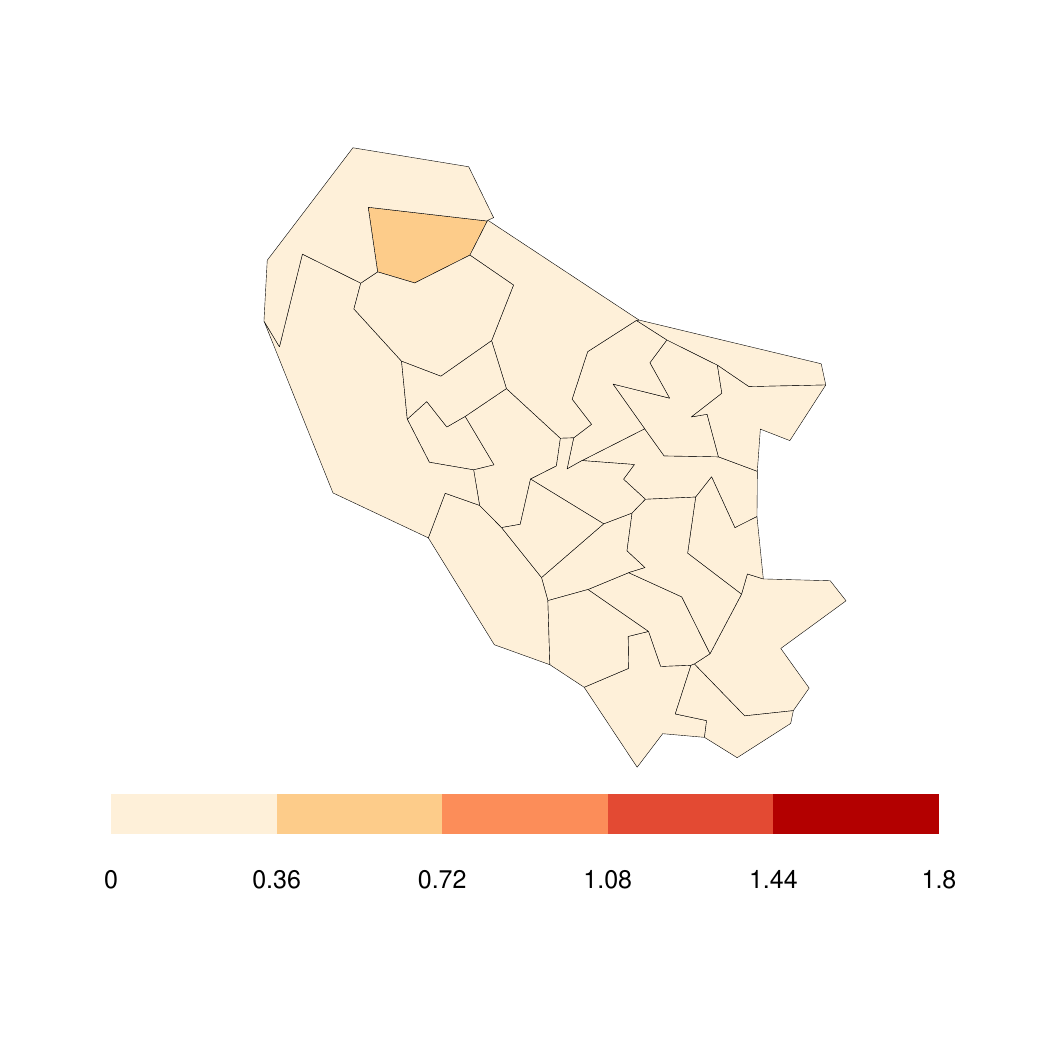}
    \includegraphics[width = 0.48\textwidth]{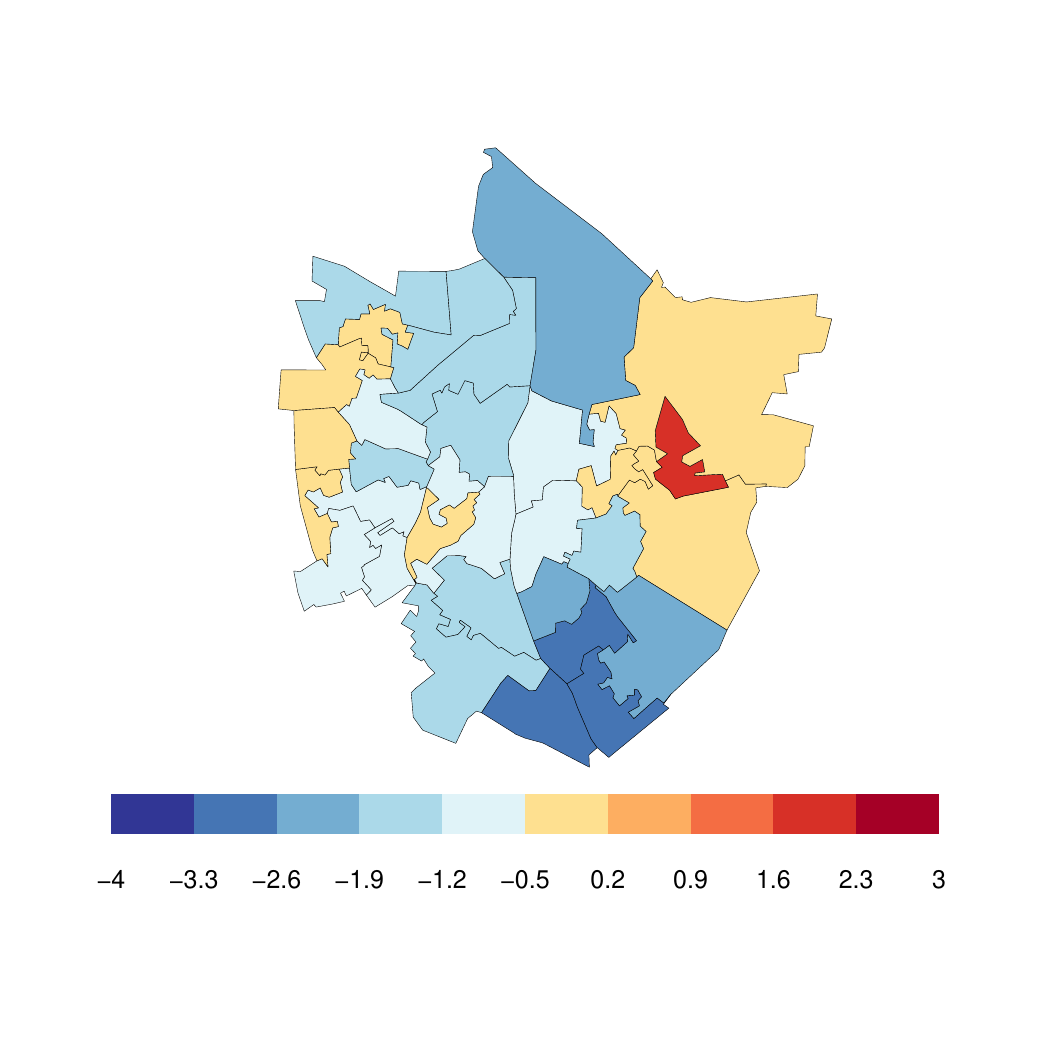}
    \includegraphics[width = 0.48\textwidth]{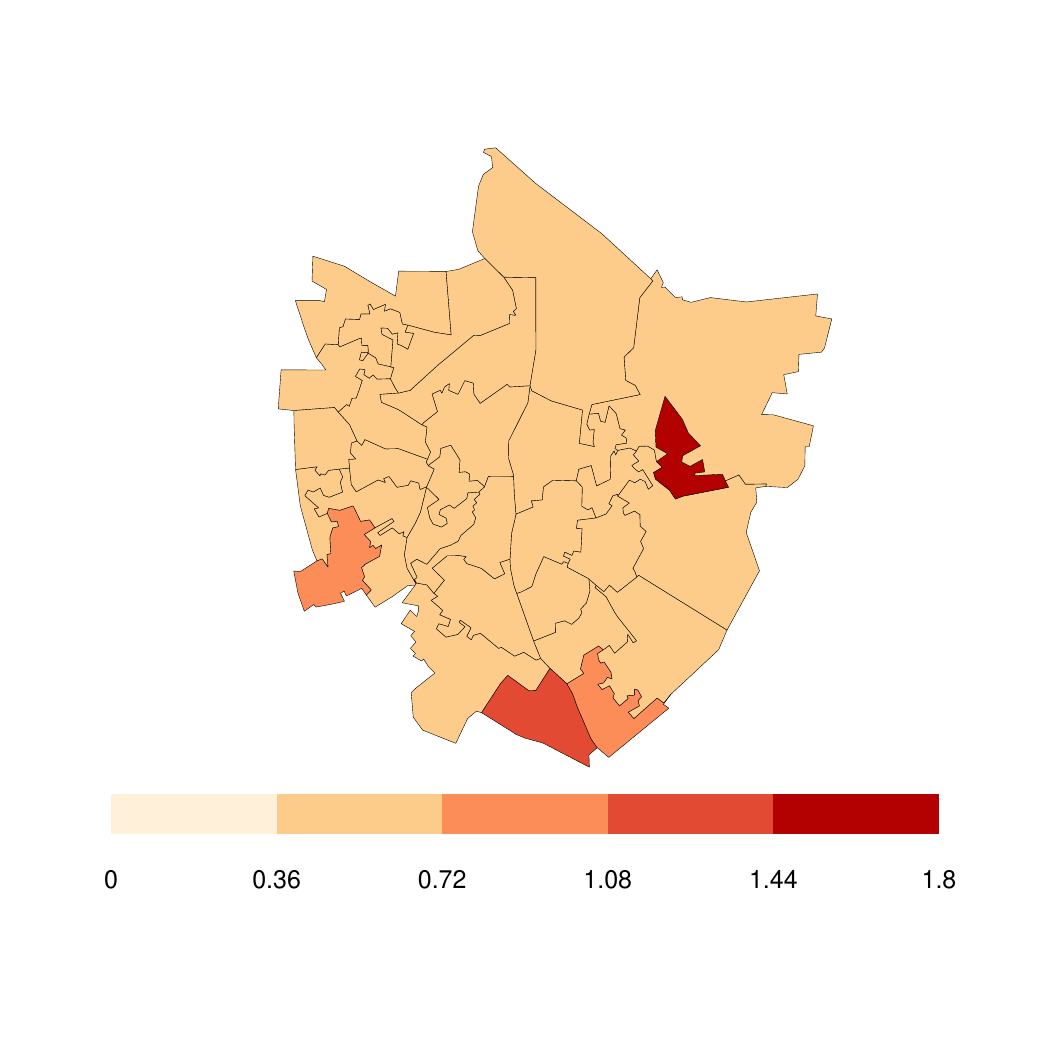}
    \caption{Top: Posterior median estimates and variance for the relative risk of honour based abuse in Oxford. Bottom: Posterior median estimates and variance for the relative risk of honour based abuse in Banbury.}
    \label{fig: Oxon Results}
\end{figure}

The posterior median estimates for risk of FGM in each ward and the values of the posterior variance are shown in Figure \ref{fig: Oxon Results}. Oxford shows a clear spatial trend, with wards in the south east of the city having the highest risk of honour based abuse. In Banbury, the vast majority of wards are estimated to have parameter values near 0, which is the average across all wards in Oxford and Banbury. There are 9 higher risk wards on the eastern and western outskirts of the town. One ward in Banbury, East Grimsbury, is estimated to have the highest risk of honour based abuse out of all the wards in Oxford and Banbury. The parameter value for East Grmisbury is particularly extreme. This is because the ward only featured in three comparisons, and each time it was chosen as the higher risk ward. This means the model assigned high uncertainty to this parameter. Had we used independent prior distributions, the estimated would have been more extreme. Overall, we received more comparisons about wards in Oxford than Banbury, leading to lower uncertainty in our estimates for Oxford. 

\begin{figure}
    \centering
    \includegraphics[width = 0.75\textwidth]{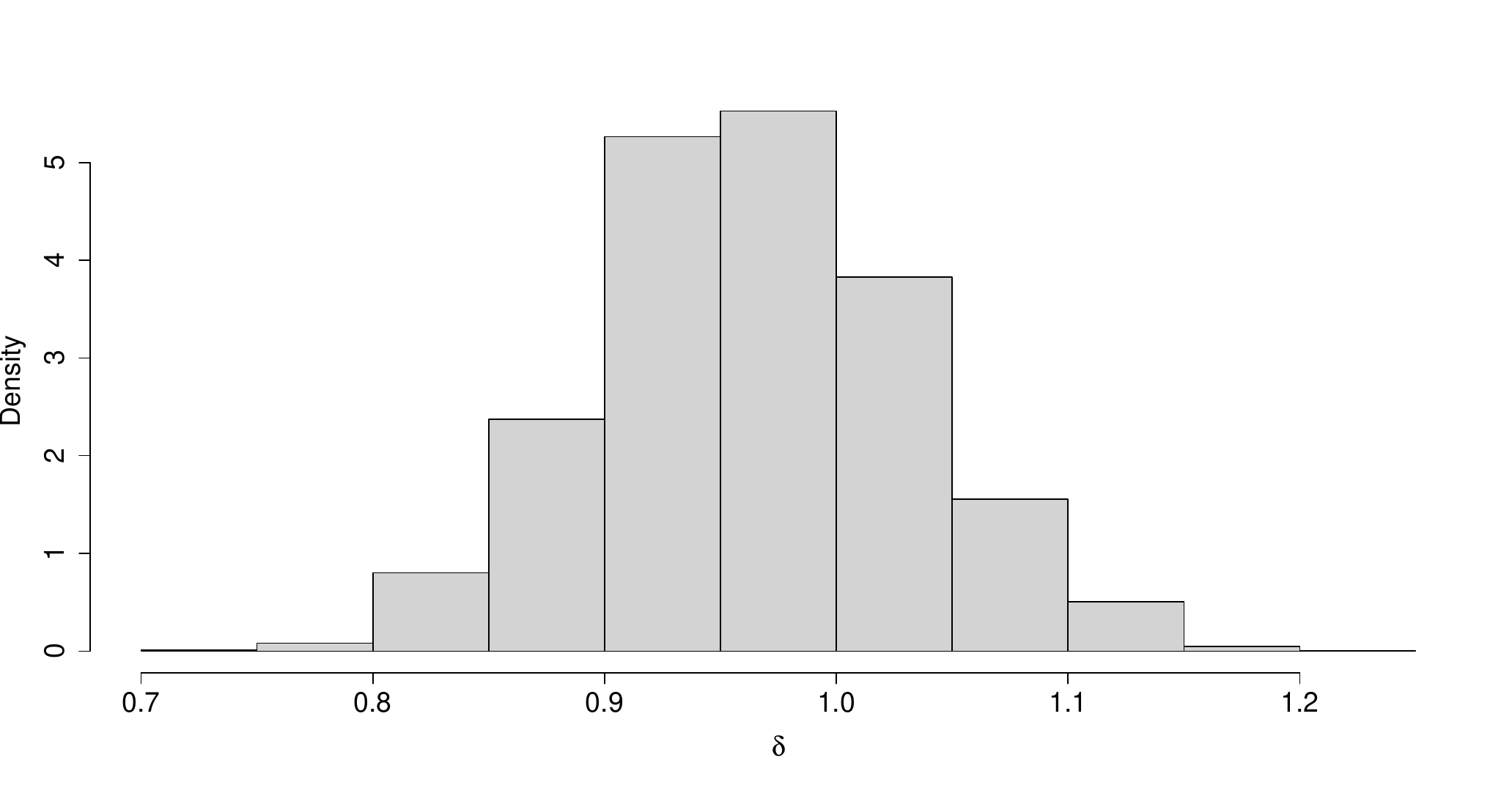}
      \includegraphics[width = 0.75\textwidth]{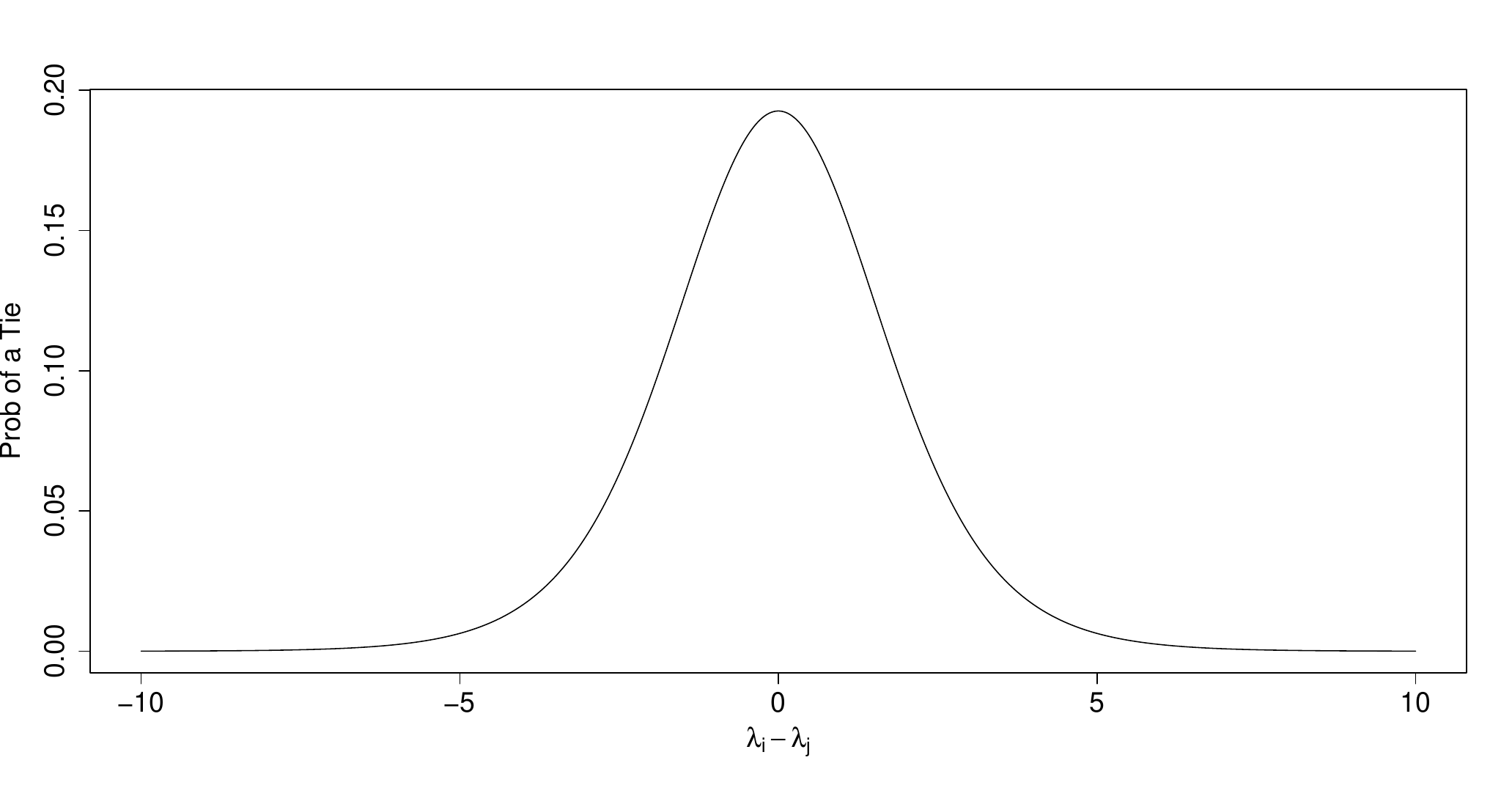}
    \caption{Left: A histogram for the posterior distribution of the tie parameter $\delta$ in the Oxfordshire study. Right: The probability of a judge declaring a tie using the posterior median value of $\delta$. }
    \label{fig: prob of a tie}
\end{figure}

We can also investigate the behaviours associated with tied comparisons. The posterior median estimate for $\delta$ is 0.964. Figure \ref{fig: prob of a tie} shows the probability of a tie against the difference in values honour based abuse. For pairs of ward with identical levels of honour based abuse, we estimate that a judge would say that the wards are tied with probability 0.193. 

\section{Conclusion} \label{sec: conclusion}
In this paper, we have developed an efficient and scalable inference method for a Bradley--Terry model with ties. This method allows for a correlated prior distribution to be placed on the model quality parameters, making it suitable for applications where there is a spatial element or little data is available. It also allows us to chose the prior distribution for the tied parameter, offering a more flexible framework than previous scalable methods.

We used our method on two new data sets about honour based abuse in UK counties, providing community level information this crime for the first time in the counties. Allowing for tied comparisons, allowed the judges to provide more accurate information about the risk of this crime. We designed a bespoke interface to collect this data and allow for tied judgement to be made. We collected over 750 comparisons in each county. Our results being used to inform community based interventions in the counties. For each study, we produced a two-page non-technical briefing describing the results (see Supplementary Material), and shared these with our project partners, local councillors and Members of the UK Parliament. Two questions were tabled in the House of Commons about Honour Based Abuse based as a result of this work -- one to the Secretary of State for Health and Social Care \cite{Q2} and one to the Home Secretary \cite{Q1}. 

There are several directions this work could be advanced. Firstly, in terms of the statistical methodology, there is a lack of theory about the number of comparisons that need to be collected to provide rigorous results. One of the limitations of experimental design is the high computational cost and the need to fit the model to many simulated data sets. As our method is efficient, it may allow for experimental design in comparative judgement to be advanced. Secondly, in terms of the application, a model including socio-demographic factors could be developed to understand drivers of Honour-Base Abuse. Thirdly, bias may be an issue in the data as we are collecting what amounts to anecdotal data. Work on goodness-of-fit and misfit detection could identify judges who have significantly different opions to the other judges about where abuse is happening. Finally, in terms of the interface, an interface could be developed to allow for online inference to be carried out. 

\section*{Acknowledgements}
We thank South Yorkshire Police and Oxford Against Cutting for their support with data collection. Digital Research Support was provided by the Birmingham Environment for Academic Research. This work was supported by a UK Research and Innovation Future Leaders Fellowship [MR/X034992/1]. We thank the two anonymous reviewers and the Associate Editor for their constructive comments. 

\section*{Data Availability}
We have developed an R package to fit models to the data. The  \texttt{speedyBBT} R package is available at CRAN \citep{SpeedyBBT} and also contains the data. The python interface we developed to collect the data is available at \url{https://gitlab.bham.ac.uk/seymourg-bsbt/comparison-interface} \cite{Smith24}. 

\bibliographystyle{tfs}
\bibliography{bibliography}

\end{document}